\documentclass{aa}

\usepackage{graphicx}
\usepackage{txfonts}

\usepackage{color}
\usepackage{bm}

\newcommand{\tred}[1]{#1}

\begin{document}

\title{
Semi-conservative reduced speed of sound technique \\
for low Mach number flows with large density variations
}

\titlerunning{New Reduced Speed of Sound Technique}

\author{
H. Iijima\inst{1}
\and
H. Hotta\inst{2}
\and
S. Imada\inst{1}
}

\institute{
Division for Integrated Studies,
Institute for Space-Earth Environmental Research, Nagoya University,
Furocho, Chikusa-ku, Nagoya, Aichi 464-8601, Japan\\
\email{h.iijima@isee.nagoya-u.ac.jp}
\and
Department of Physics, Faculty of Science, Chiba University,
1-33 Yayoi-chou, Inage-ku, Chiba 263-8522, Japan
}

\date{Received; accepted}

\abstract
{
The reduced speed of sound technique (RSST)
has been used for efficient simulation
of low Mach number flows
in solar and stellar convection zones.
The basic RSST equations are hyperbolic,
and are suitable for parallel computation
by domain decomposition.
The application of RSST is limited
to cases where density perturbations
are much smaller than the background density.
In addition, non-conservative variables
are required to be evolved using this method,
which is not suitable in cases where 
discontinuities like shock waves
co-exist in a single numerical domain.
}
{
In this study, we suggest a
new semi-conservative formulation of the RSST
that can be applied to low Mach number flows
with large density variations.
}
{
We derive the wave speed
of the original and newly suggested methods
to clarify that these methods can
reduce the speed of sound without
affecting the entropy wave.
The equations are implemented
using the finite volume method.
\tred{
Several numerical tests are carried out
to verify the suggested methods.
}
}
{
The analysis and numerical results
show that the original RSST is not applicable
when mass density variations are large.
In contrast, the newly suggested methods
are found to be efficient in such cases.
We also suggest variants of the RSST
that conserve momentum in the machine precision.
The newly suggested variants
are formulated as semi-conservative equations,
which reduce to the conservative form
of the Euler equations when the speed of sound is not reduced.
This property is advantageous when
both high and low Mach number regions are
included in the numerical domain.
}
{
The newly suggested forms of RSST
can be applied to a wider range of low Mach number flows.
}


\maketitle

\newpage
\section{Introduction}

Low Mach number flows sometimes appear
in astrophysical phenomena.
One typical example is the stellar convection zone,
where the Mach number is very small ($\sim10^{-3}$ or less).
In such cases, the time step criterion for explicit methods
is severely limited by the fast speed of sound via the CFL condition,
even if we are interested in the much slower dynamics
of convective motion.
Various numerical methods are suggested
for efficient simulation of the low Mach number flows
\citep[see][for a review of the numerical modeling of the stellar convection]
{2017LRCA....3....1K}.

One major way to simulate low Mach number flows
is to assume that the speed of sound is infinite,
as is done in the Boussinesq/anelastic approximations
\citep[e.g.,][]{2008ApJ...673..557M}.
The drawback of the these methods
is the necessity of global communication
in parallel computing.
A sound wave with infinite speed forces the entire computational domain
to interact instantaneously.
Implicit time integration
of the sound wave as in the stratified approximation
\citep{1994JCoPh.113..165C,2016JCoPh.310..342C}
also suffers from this difficulty.
This characteristic can be a source
that prevents high parallel computing efficiency
in a numerical simulation.

The reduce speed of sound technique (RSST) was 
first developed to compute the steady state solution
\citep{2005ApJ...622.1320R}
and was extended to the thermal convection problem
\citep{2012A&A...539A..30H}.
In the RSST, speed of sound is artificially reduced
by a free parameter $\xi$
so that the severe time step criterion
due to a fast speed of sound can be relaxed.
The equations are fully explicit and can be easily implemented
with parallel computers using domain decomposition.
\cite{2015ApJ...798...51H} also suggests
an invariant that exhibits better conservation properties.
The RSST has been previously applied to solar and stellar
convection problems with and without magnetic fields
\citep{2014ApJ...786...24H,2015ApJ...798...51H,
2016Sci...351.1427H,2016A&A...588A.150K}.
In contrast to the high Reynolds number
in stellar convection,
mantle convection in the earth has
very low Reynolds number.
\cite{2017NewA...50...82T} proposed a method to solve such problems
using a strategy that is similar to RSST.
In the following, we concentrate on
applications to high Reynolds number flows.

One limitation of the original RSST
is that it was formulated and tested
for problems where density perturbations
are sufficiently smaller than the background density.
As shown in Sec. \ref{sec:dvs} and \ref{sec:test},
we find that the original version of the RSST cannot be used to handle
low Mach number flows with large density variations.
Large density variations in low Mach number flows
sometimes appear in problems
with the state equation for non-ideal gases
and/or nuclear or chemical reactions
like the convective phase of Type Ia Supernova
\citep{2005ApJ...632.1021Z,2007ApJ...665.1321G,
2010ApJS..188..358N}.

Another limitation of the original RSST
is that the evolution equations
for non-conservative variables
(velocity field and specific entropy)
must be solved.
This limitation comes from the requirement
that the evolution of the entropy equation be solved directly
so that artificial reduction of the speed of sound
does not affect the characteristics of the entropy wave.
However, the conservative form is preferred
when the solution includes discontinuities like shock waves,
especially in finite-difference and finite-volume schemes
\citep{1994MaCom..62..497H}.
Let us take an example from stellar convection simulations,
where the numerical domain extends
from the deep stellar convection zone
to the photosphere or upper atmosphere.
The basic RSST equations in the deep convection zone
do not reduce smoothly to the conservative equations
used in surface convection simulations
\citep{1982A&A...107....1N,1989ApJ...342L..95S,
2005A&A...429..335V,2015ApJ...812L..30I,2017ApJ...848...38I}.

In this study, we suggest several new formulations
to reduce the speed of sound
while maintaining the properties of the entropy wave.
The designed methods have two advantages:
(1) It can be applied to flows with large density variation.
(2) It evolves the conservative variables
and reduces to the conservative Euler equations
when the speed of sound is not reduced.
These advantages of the newly suggested RSST methods
lead to their application in a wide range of problems.

\section{Original formulation
 of the reduced speed of sound technique (DVS form)\label{sec:dvs}}

In this section, we summarize the original formulation
of the RSST \citep{2005ApJ...622.1320R,2012A&A...539A..30H}
and show that this method cannot be applied to problems
with large density variations.
In this study, we call this original version of the RSST
the DVS form (each abbreviation shows
D: density, V: velocity, and S: entropy).

\subsection{Reducing the time derivative of mass density
  \label{subsec:dvs_prim}}

The original formulation of the RSST
in Cartesian coordinates is formulated
by dividing the background stratification by perturbations in the system.
The inviscid Euler equations with the RSST is given by
\begin{equation}
 \label{eq:rsst_zero}
 \begin{aligned}
  \xi^2\frac{\partial{\rho_1}}{\partial{t}}
  +\nabla\cdot\left(\rho_0\bm{V}\right)=0\\
  \frac{\partial{V_i}}{\partial{t}}+\bm{V}\cdot\nabla{V_i}
  +\frac{1}{\rho_0}\frac{\partial{P_1}}{\partial{x_i}}=0\\
  \frac{\partial{S_1}}{\partial{t}}
  +\bm{V}\cdot\nabla\left(S_0+S_1\right)=0
 \end{aligned}
\end{equation}
where $\rho$ is the mass density,
$\bm{V}$ is the velocity field,
$P$ is the gas pressure,
$S$ is the specific entropy,
and $i=x,y,z$.
The subscripts $0$ and $1$ denote
the background stratification and perturbation, respectively.
Here, the perturbation is assumed to be small
and the background stratification does not vary in time
\begin{align}
 \label{eq:zero_assum}
 P_1\ll{P_0},\rho_1\ll{\rho_0},\
 \partial_t{P_0}=\partial_t{\rho_0}=0
\end{align}
The important point is that
the entropy equations are not altered by applying the RSST.
This allows us to evolve flows without
modifying the characteristics of the entropy wave.
We assume that $P_0$ is spatially uniform in the above formulation.
In a practical situation, the gradient of $P_0$
balances the external forces (e.g., the gravitational force).
This difference does not affect the discussion in this paper.

We rewrite the original equations (Eqs. (\ref{eq:rsst_zero}))
without dividing by the background and perturbations in the variables,
without changing the phase speed of each mode 
under the limit in Eq. (\ref{eq:zero_assum}).
The resulting equations are given by
\begin{equation}
 \label{eq:dvs_prim}
  \begin{aligned}
   \xi^2\frac{\partial{\rho}}{\partial{t}}
   +\nabla\cdot\left(\rho\bm{V}\right)=0\\
   \frac{\partial{V_i}}{\partial{t}}+\bm{V}\cdot\nabla{V_i}
   +\frac{1}{\rho}\frac{\partial{P}}{\partial{x_i}}=0\\
   \frac{\partial{S}}{\partial{t}}+\bm{V}\cdot\nabla{S}=0
  \end{aligned}
\end{equation}
where we assume the general equation of state $P(\rho,S)$.

Let us consider a plane wave in the $x$-direction
and analyze the phase speed of each wave mode.
The one-dimensional version
of Eqs. (\ref{eq:dvs_prim}) is given by
\begin{equation}
\begin{aligned}
 \xi^2\frac{\partial{\rho}}{\partial{t}}
 +V_x\frac{\partial{\rho}}{\partial{x}}
 +\rho\frac{\partial{V_x}}{\partial{x}}=0\\
 \frac{\partial{V_x}}{\partial{t}}
 +V_x\frac{\partial{V_x}}{\partial{x}}
 +\frac{1}{\rho}\frac{\partial{P}}{\partial{x}}=0\\
 \frac{\partial{S}}{\partial{t}}
 +V_x\frac{\partial{S}}{\partial{x}}=0
\end{aligned}
\end{equation}
These equations can be rewritten as
\begin{equation}
\begin{aligned}
 &\frac{\partial}{\partial{t}}
 \begin{pmatrix}
  \rho\\
  V_x\\
  S
 \end{pmatrix}
 +A\frac{\partial}{\partial{x}}
 \begin{pmatrix}
  \rho\\
  V_x\\
  S
 \end{pmatrix}
 =0\\
 A&=
 \begin{pmatrix}
  1/\xi^2 & 0 & 0\\
  0 & 1 & 0\\
  0 & 0 & 1\\
 \end{pmatrix}
 \begin{pmatrix}
  V_x & \rho & 0\\
  a^2/\rho & V_x & P_S/\rho\\
  0 & 0 & V_x\\
 \end{pmatrix}\\
 &=
 \begin{pmatrix}
  V_x/\xi^2 & \rho/\xi^2 & 0\\
  a^2/\rho & V_x & P_S/\rho\\
  0 & 0 & V_x\\
 \end{pmatrix}
\end{aligned}
\end{equation}
where $P_S=(\partial{P}/\partial{S})_\rho$,
and $a=\sqrt{(\partial{P}/\partial{\rho})_S}$
is the adiabatic speed of sound.
In the case of an ideal gas,
$(\partial{P}/\partial{\rho})_S=\gamma{P/\rho}$
and $(\partial{P}/\partial{S})_\rho=P/C_V$,
where $\gamma$ is the specific heat ratio
and $C_V$ is the specific heat capacity
at constant volume.

The phase speed of each wave mode
is equal to an eigenvalue of the Jacobian $A$,
which is given by
\begin{align}
 \label{eq:dvs_lambda1}
 \lambda=
 \begin{cases}
  V_x,\\
  \frac{1}{2}\left[
  \left(1+\frac{1}{\xi^2}\right)V_x\pm\sqrt{D}
  \right]
 \end{cases}
\end{align}
where
\begin{align}
 \label{eq:dvs_lambda2}
 D=\left(1-\frac{1}{\xi^2}\right)^2V_x^2+4\frac{a^2}{\xi^2}
\end{align}
Apparently, $D>0$ is always satisfied
and all eigenvalues are real,
which indicates that the basic RSST equations
in the DVS form (Eqs. (\ref{eq:dvs_prim}))
are hyperbolic.
The first wave mode $\lambda=V_x$
corresponds to the entropy wave.
The other two wave modes
correspond to right/left-propagating sound waves.
When the Mach number is small enough ($|\bm{V}|\ll{C}$),
the speed of the sound wave
becomes $\lambda\sim\pm{a/\xi}$.
Thus, the RSST can be used to reduce the speed of sound
without affecting the characteristics of the entropy wave.

\subsection{Drawback from reducing
  the time derivative of mass density\label{subsec:dvs_drawback}}

Although the original DVS form
is simple and easy to implement,
this approximation
excites an artificial pressure perturbation
when the density perturbation
is not small compared to the background.
From the RSST continuity and entropy equations
in Eqs. (\ref{eq:dvs_prim}),
we find that the equation describing evolution of the gas pressure is given by
\begin{align}
 \label{eq:dvs_pr}
 \frac{\partial{P}}{\partial{t}}
 +\bm{V}\cdot\nabla{P}
 +\rho{a^2}\nabla\cdot\bm{V}=
 \left(1-\frac{1}{\xi^2}\right)a^2
 \left(
 \bm{V}\cdot\nabla{\rho}
 +{\rho}\nabla\cdot\bm{V}
 \right)
\end{align}
Let us assume a situation
where the velocity field is
incompressible (i.e., $\nabla\cdot\bm{V}=0$)
and the gas pressure is uniform (i.e., $\nabla{P}=0$).
Such a situation nearly occurs during the evolution
of low Mach number flows,
because the gas pressure is adjusted
by the fast sound wave
until the forces balance each other.
Eq. (\ref{eq:dvs_pr}) leads to
\begin{align}
 \frac{\partial{P}}{\partial{t}}=
 \left(1-\frac{1}{\xi^2}\right)a^2
 \left(
 \bm{V}\cdot\nabla{\rho}
 \right)
\end{align}
This means that density advection
causes a pressure imbalance
and excites artificial sound waves.
This drawback is clearly shown in a Kelvin-Helmholtz instability test problem (Sec. \ref{subsec:kh01}).
If the RSST is not used (i.e., $\xi=1$),
then pressure balance is maintained.

The RSST has been applied to the deep region
of the stellar convection zone,
where mass density perturbations
are very small compared to the background density ($\sim10^{-4}$ or less).
In such a case, the amplitude of
the artificially excited pressure perturbation is negligible.
However, the RSST would not be applicable
to a situation with large density variations,
which occurs in some astrophysical phenomena.

\section{Reduced speed of sound technique
for problems with large density variation (PVS form)\label{sec:pvs}}

In this section,
we propose a new RSST formulation named ``PVS form''
(P: pressure, V: velocity, and S: entropy)
that is suitable for low Mach number flows
with large density variations.
In Sec. \ref{subsec:pvs_prim},
we introduce the basic PVS form equations,
and we investigate their characteristics.
In Sec. \ref{subsec:pvs_cons},
we show that the PVS form can be rewritten
in a semi-conservative form
so that conservative variables can be evolved directly.
This is advantageous because the basic equations of the RSST
becomes the conservative Euler equations when the speed of sound is not reduced ($\xi=1$),
which is suitable for solving systems that contain discontinuous phenomena
like shock waves.

\subsection{Reducing the temporal evolution
  of gas pressure\label{subsec:pvs_prim}}

The PVS form is a new formulation of the RSST
based on equations describing the evolution of $(P,\bm{V},S)$.
In this formulation,
we limit the time variation of the gas pressure
instead of the mass density to reduce the speed of sound.
The basic equations are given by
\begin{equation}
 \label{eq:pvs_prim}
  \begin{aligned}
 \xi^2\frac{\partial{P}}{\partial{t}}
 +\bm{V}\cdot\nabla{P}
 +\rho{a^2}\nabla\cdot\bm{V}=0\\
 \frac{\partial{V_i}}{\partial{t}}+\bm{V}\cdot\nabla{V_i}
 +\frac{1}{\rho}\frac{\partial{P}}{\partial{x_i}}=0\\
 \frac{\partial{S}}{\partial{t}}+\bm{V}\cdot\nabla{S}=0
  \end{aligned}
\end{equation}

Apparently, the new formulation in Eqs. (\ref{eq:pvs_prim})
maintains pressure equilibrium
when the initial pressure is uniform
and the divergence of the velocity field is zero.
This allows
the time variation of the pressure
to be reduced instead of reducing the density variation.
We find this characteristic to be advantageous
in more practical situations, such as those tested in Sec. \ref{sec:test}.

Eqs. (\ref{eq:pvs_prim}) have wave speeds that are identical to the wave speeds in the original DVS form.
The one-dimensional version of Eqs. (\ref{eq:pvs_prim})
is given by
\begin{align}
 &\frac{\partial}{\partial{t}}
 \begin{pmatrix}
  P\\
  V_x\\
  S
 \end{pmatrix}
 +A\frac{\partial}{\partial{x}}
 \begin{pmatrix}
  P\\
  V_x\\
  S
 \end{pmatrix}
 =0\\
 A&=
 \begin{pmatrix}
  V_x/\xi^2 & \rho{a^2}/\xi^2 & 0\\
  1/\rho & V_x & 0\\
  0 & 0 & V_x\\
 \end{pmatrix}
\end{align}
The eigenvalues of the Jacobian $A$ are given by
\begin{align}
 &\lambda=
 \begin{cases}
  V_x,\\
  \frac{1}{2}\left[
  \left(1+\frac{1}{\xi^2}\right)V_x\pm\sqrt{D}
  \right]
 \end{cases}\\
 &D=\left(1-\frac{1}{\xi^2}\right)^2V_x^2+4\frac{a^2}{\xi^2}
\end{align}
The wave speeds $\lambda$ are identical to the speeds in
Eqs. (\ref{eq:dvs_lambda1}) and (\ref{eq:dvs_lambda2}).
The new PVS form can also reduce the speed of sound
without affecting the nature of the entropy wave, as is the case in the DVS form.

\subsection{PVS form for conservative variables\label{subsec:pvs_cons}}

The basic PVS form equations
(Eqs. \ref{eq:pvs_prim})
can be rewritten in the semi-conservative form as
\begin{equation}
 \label{eq:pvs_cons}
  \begin{aligned}
 \frac{\partial{\rho}}{\partial{t}}&
 +\nabla\cdot\left(\rho\bm{V}\right)=
 -\left(1-\frac{1}{\xi^2}\right)
 \left(
 \frac{\partial{\rho}}{\partial{P}}
 \right)_{S}\varDelta{P}
 \\
 \frac{\partial}{\partial{t}}\left({\rho{V_i}}\right)&
 +\nabla\cdot\left(\rho{V_i}\bm{V}\right)
 +\frac{\partial{P}}{\partial{x_i}}=
 -\left(1-\frac{1}{\xi^2}\right)
 \left(
 \frac{\partial{\rho{V_i}}}{\partial{P}}
 \right)_{\bm{V},S}\varDelta{P}
 \\
 \frac{\partial{E}}{\partial{t}}&
 +\nabla\cdot\left[\left(E+P\right)\bm{V}\right]
 =
 -\left(1-\frac{1}{\xi^2}\right)
 \left(
 \frac{\partial{E}}{\partial{P}}
 \right)_{\bm{V},S}\varDelta{P}
  \end{aligned}
\end{equation}
where $E=e+\frac{1}{2}\rho{V^2}$ is the total energy density,
and $e(\rho,P)$ is the internal energy density
(for the ideal equation of state, $e=P/(\gamma-1)$).
The partial derivatives on the right hand side
of Eqs. (\ref{eq:pvs_cons}) are given by
\begin{align}
 \label{eq:pvs_cons_rhs}
 \left(
 \frac{\partial{\rho}}{\partial{P}}
 \right)_{S}
 =\frac{1}{a^2},
 \
 \left(
 \frac{\partial{\rho{V_i}}}{\partial{P}}
 \right)_{\bm{V},S}
 =\frac{V_i}{a^2},
 \
 \left(
 \frac{\partial{E}}{\partial{P}}
 \right)_{\bm{V},S}
 =\frac{e+P+\rho{V^2}/2}{\rho{a^2}}
\end{align}
$\varDelta{P}$ is the time derivative
of the gas pressure before reducing the speed of sound
and is given by
\begin{equation}
 \label{eq:pvs_cons_dpr}
  \begin{aligned}
 \varDelta{P}
 &=
 -\bm{V}\cdot\nabla{P}
 -\rho{a^2}\nabla\cdot\bm{V}
 \\
 &=
 -\left[\left(
 \frac{\partial{e}}{\partial{P}}
 \right)_{\rho}\right]^{-1}
 \left\{
 \left[
 \frac{1}{2}V^2
 -\left(\frac{\partial{e}}{\partial{\rho}}
 \right)_{P}\right]
 \nabla\cdot\left(\rho\bm{V}\right)
 \right.
 \\
 &\hspace{1cm}\left.
 -V_i\left[\nabla\cdot\left(\rho{V_i}\bm{V}\right)
 +\frac{\partial{P}}{\partial{x_i}}\right]
 +\nabla\cdot\left[\left(E+P\right)\bm{V}\right]
 \right\}
  \end{aligned}
\end{equation}
The thermodynamic derivatives are given by
\begin{align}
 \label{eq:pvs_cons_therm}
 \left(\frac{\partial{e}}{\partial{P}}\right)_{\rho}
 =\frac{\rho{T}}{P_S}
 ,\
 \left(\frac{\partial{e}}{\partial{\rho}}\right)_{P}
 =\frac{e+P}{\rho}-\frac{\rho{Ta^2}}{P_S}
\end{align}
In the state equation for an ideal gas,
$\left(\partial{e}/\partial{P}\right)_\rho=1/\left(\gamma-1\right)$
and $\left(\partial{e}/\partial{\rho}\right)_P=0$.
Eqs. (\ref{eq:pvs_cons}) indicate
that the speed of sound can be reduced
as it was in the the PVS form
by adding a correction term
to the right hand side of the conservative Euler equations.
This reduces the pressure variation while ensuring that variations in the specific entropy
and velocity field are unchanged.

By solving Eqs. (\ref{eq:pvs_cons}),
we can reduce the speed of sound
without affecting the entropy wave.
When the speed of sound is not reduced (i.e., $\xi=1$),
the equations reduce to the conservative Euler equations.
Thus, the equations can be easily applied
to problems with both shock waves and low Mach number flows.

Introducing new temporary variables
\begin{equation}
 \label{eq:pvs_dt0}
  \begin{aligned}
 \varDelta{\rho}
 &=-\nabla\cdot\left(\rho\bm{V}\right)\\
 \varDelta\left(\rho{V_i}\right)
 &=-\nabla\cdot\left(\rho{V_i}\bm{V}\right)
 -\frac{\partial{P}}{\partial{x_i}}\\
 \varDelta{E}
 &=
 -\nabla\cdot\left[\left(E+P\right)\bm{V}\right]
  \end{aligned}
\end{equation}
Eqs. (\ref{eq:pvs_cons})--(\ref{eq:pvs_cons_therm})
can be rewritten as
\begin{equation}
 \label{eq:pvs_dt}
  \begin{aligned}
 \frac{\partial{\rho}}{\partial{t}}
 &=\varDelta{\rho}
 -\left(1-\frac{1}{\xi^2}\right)
 \left(
 \frac{\partial{\rho}}{\partial{P}}
 \right)_{S}\varDelta{P}
 \\
 \frac{\partial}{\partial{t}}\left({\rho{V_i}}\right)
 &=\varDelta\left(\rho{V_i}\right)
 -\left(1-\frac{1}{\xi^2}\right)
 \left(
 \frac{\partial{\rho{V_i}}}{\partial{P}}
 \right)_{\bm{V},S}\varDelta{P}
 \\
 \frac{\partial{E}}{\partial{t}}
 &=\varDelta{E}
 -\left(1-\frac{1}{\xi^2}\right)
 \left(
 \frac{\partial{E}}{\partial{P}}
 \right)_{\bm{V},S}\varDelta{P}
  \end{aligned}
\end{equation}
where
\begin{align}
 \label{eq:pvs_dt_dpr}
 \varDelta{P}
 =
 \left[\left(
 \frac{\partial{e}}{\partial{P}}
 \right)_{\rho}\right]^{-1}
 \left\{
 \left[
 \frac{1}{2}V^2
 -\left(\frac{\partial{e}}{\partial{\rho}}
 \right)_{P}\right]
 \varDelta{\rho}
 -V_i
 \varDelta\left(\rho{V_i}\right)
 +\varDelta{E}
 \right\}
\end{align}

Eqs. (\ref{eq:pvs_dt0})--(\ref{eq:pvs_dt_dpr})
indicate that the proposed method
can be easily implemented in existing numerical solvers
for the conservative Euler equations
by changing only the time derivatives
(see also our implementation in Sec. \ref{subsec:num_method}).
The additional computation is local
and is suitable for use with the domain decomposition technique
on parallel computers.

The steady solution ($\partial/\partial{t}=0$)
of the semi-conservative form (Eqs. (\ref{eq:pvs_cons}))
is identical to the original Euler equations.
This characteristic is apparent from
Eqs. (\ref{eq:pvs_dt0})--(\ref{eq:pvs_dt_dpr}).
This is advantageous not only for the steady problem,
but also for quasi-steady problems
such as thermal convection.

\section{Test problems\label{sec:test}}

\begin{figure*}[!t]
 \centering
 \includegraphics[width=0.8\hsize]{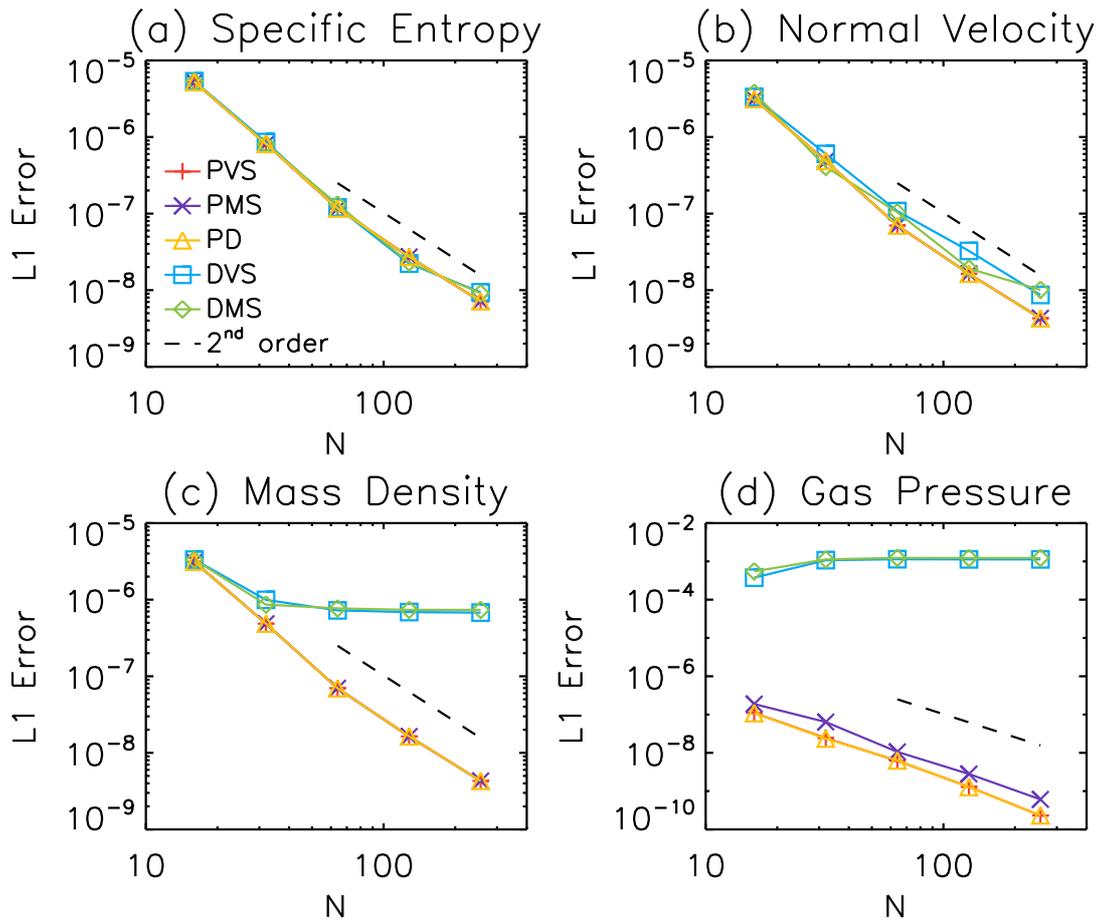}
 \caption{
 \tred{
 Linear wave convergence of entropy wave.
 The horizontal axis $N$ represents the number of grid points
 used in the $x$-direction.
 Shown are the L1 errors of (a) the specific entropy,
 (b) velocity component normal
 to the propagation direction of wave $V_\perp$,
 (c) mass density, and (d) gas pressure.
 Each line corresponds to the form of RSST used in the simulation.
 The dashed line in each panel shows
 the analytical line of second-order convergence.
 Details of the test problem are given in Sec. \ref{subsec:error}.
 }
 \label{fig:err_cmp4}
 }
\end{figure*}

\begin{figure*}[!t]
 \centering
 \includegraphics[width=0.8\hsize]{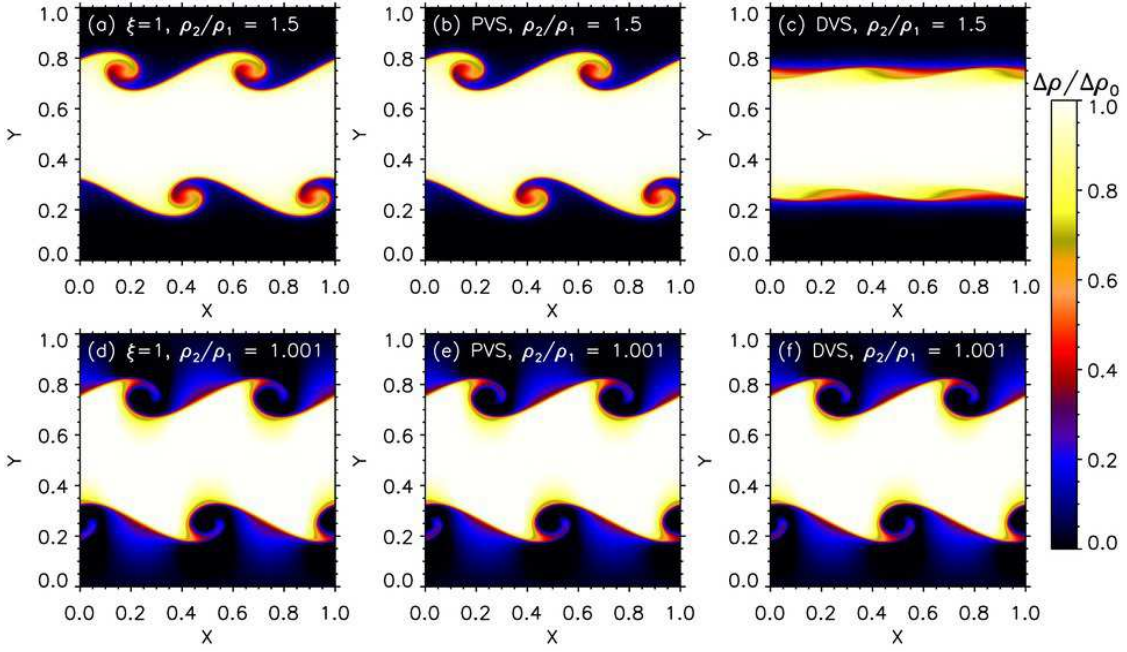}
 \caption{
 Snapshots from the Kelvin-Helmholtz instability
 with low density contrast at time $=1.5$.
 The normalized density variation
 $\varDelta\rho/\varDelta\rho_0
 =(\rho-\rho_1)/(\rho_2-\rho_1)$ is shown.
 Panels in the top row show
 results with density contrast of $1.5$.
 Panels in the bottom row show
 results with density contrast of $1.001$.
 Each column corresponds to a different
 reduction method for the speed of sound:
 without the RSST ($\xi=1$; panels (a) and (d)),
 with the PVS form ($\xi=3$; panels (b) and (e)), and
 with the DVS form ($\xi=3$; panels (c) and (f)).
 Details of the test problem are given in Sec. \ref{subsec:kh01}.
 \label{fig:kh02_f01}
 }
\end{figure*}


\begin{figure*}[!t]
 \centering
 \includegraphics[width=0.8\hsize]{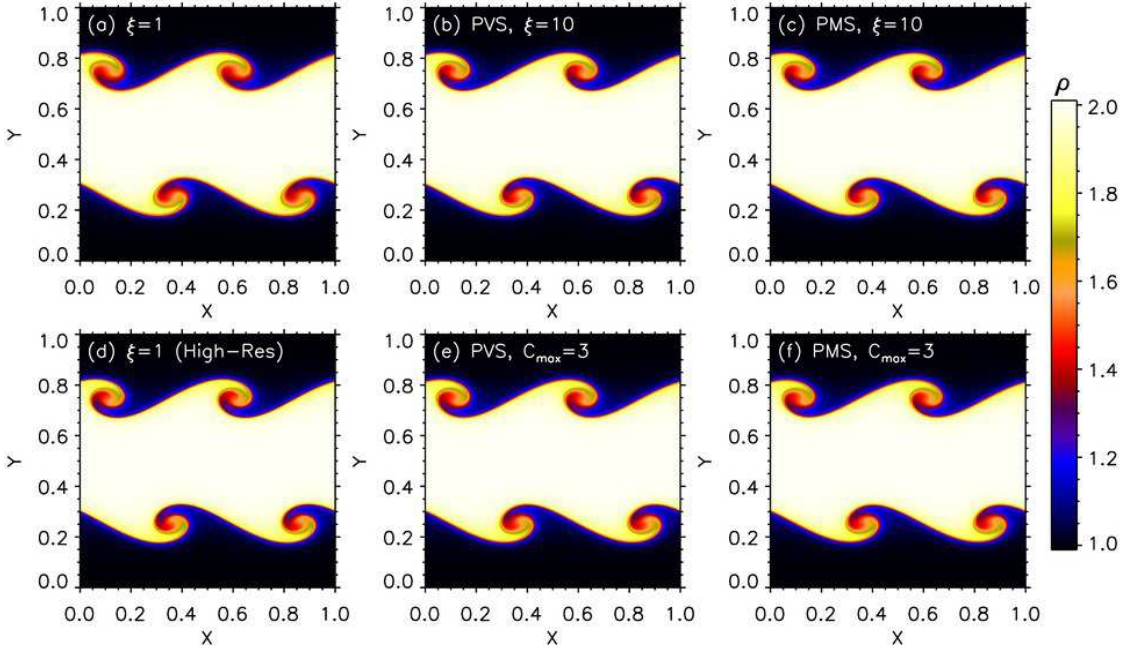}
 \caption{
 Snapshots of the Kelvin-Helmholtz instability
 with density contrast of $2$ at time $=1.5$.
 The color contour shows the mass density
 (a) without the RSST ($\xi=1$),
 (b) with the PVS form ($\xi=10$),
 (c) with the PMS form ($\xi=10$),
 (d) without the RSST ($\xi=1$) in $4096\times4096$ grids,
 (e) with the PVS form ($C_\mathrm{max}=3$), and
 (f) with the PMS form ($C_\mathrm{max}=3$).
 The detail of the test problem is shown in Sec. \ref{subsec:kh01}.
 \label{fig:kh01_f01}
 }
\end{figure*}

\begin{figure}[!t]
 \centering
 \includegraphics[width=0.8\hsize]{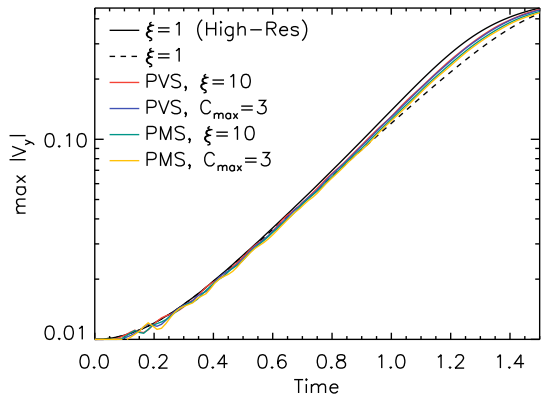}
 \caption{
 Time evolution of the Kelvin-Helmholtz instability
 with density contrast of $2$.
 The figure shows the maximum amplitude of the $y$-component of the velocity
 without RSST ($\xi=1$)
 in $4096\times4096$ grids (black solid),
 without RSST ($\xi=1$)
 in the default $1024\times1024$ grids (black dashed),
 with PVS form ($\xi=10$, red),
 with PVS form ($C_\mathrm{max}=3$, blue),
 with PMS form ($\xi=10$, green), and
 with PMS form ($C_\mathrm{max}=3$, yellow).
 \label{fig:kh01_f02}
 }
\end{figure}

\begin{figure}[!t]
 \centering
 \includegraphics[width=\hsize]{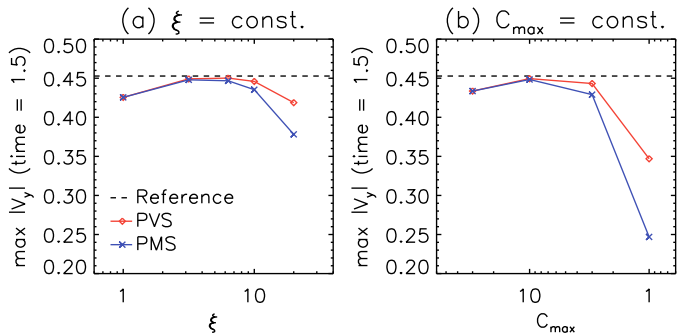}
 \caption{
 Speed of sound reduction rate
 and its dependence on the Kelvin-Helmholtz instability
 with density contrast of $2$
 with (a) $\xi=\mathrm{const.}$
 and (b) $C_\mathrm{max}=\mathrm{const}$.
 In each panel, the maximum amplitude of
 the $y$-component of the velocity at time $=1.5$
 with the PVS form (red with diamond) and
 the PMS form (blue with cross) are shown. The horizontal dashed line 
 shows the reference simulation results without the RSST ($\xi=1$)
 in $4096\times4096$ grids.
 \label{fig:kh01_f04}
 }
\end{figure}

\begin{figure}[!t]
 \centering
 \includegraphics[width=0.8\hsize]{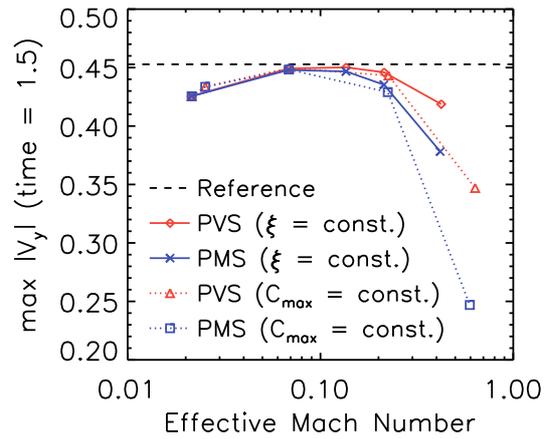}
 \caption{
 Kelvin-Helmholtz instability
 with density contrast of $2$ and its dependence on the effective Mach number.
 The figure shows the maximum amplitude of
 the $y$-component of the velocity at time $=1.5$
 in the PVS form with constant $\xi$
 (red solid with diamond),
 in the PMS form with constant
 $\xi$ (blue solid with cross),
 in the PVS form with constant $C_\mathrm{max}$
 (red dotted with triangle), and
 in the PMS form with constant $C_\mathrm{max}$
 (blue dotted with box). The horizontal dashed line 
 shows the reference simulation results without the RSST ($\xi=1$)
 in $4096\times4096$ grids.
 \label{fig:kh01_f05}
 }
\end{figure}

\begin{figure*}[!t]
 \centering
 \includegraphics[width=0.8\hsize]{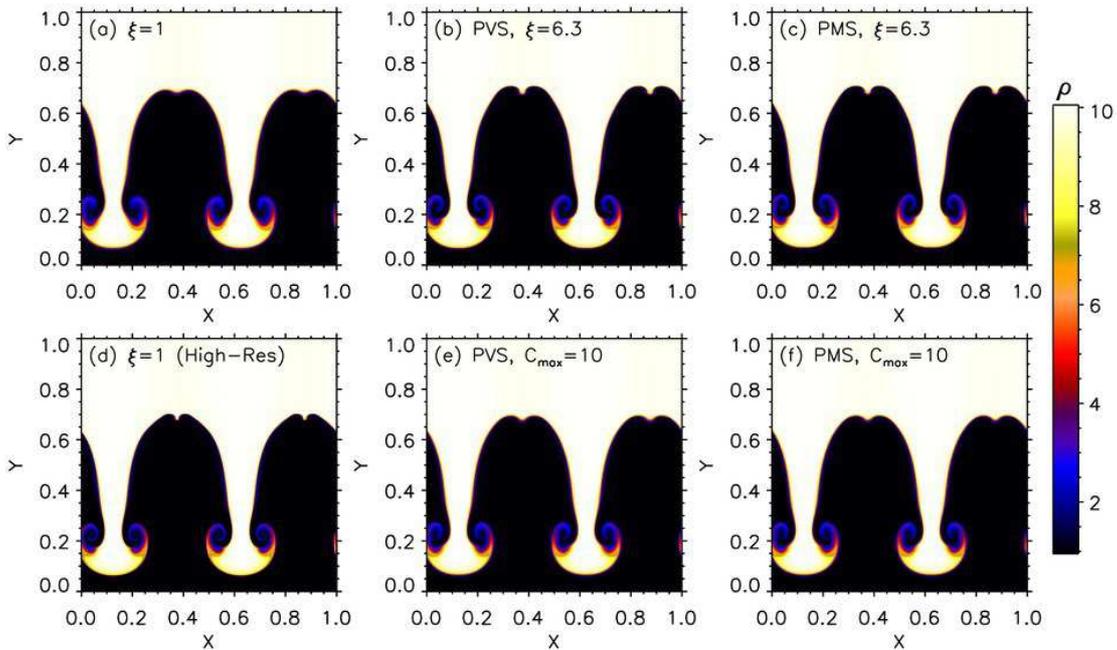}
 \caption{
 Snapshots from the Rayleigh-Taylor instability
 at time $=2.5$.
 The color contour shows the mass density
 (a) without the RSST ($\xi=1$),
 (b) with PVS form ($\xi=6.3$),
 (c) with PMS form ($\xi=6.3$),
 (d) without RSST ($\xi=1$) in $4096\times4096$ grids,
 (e) with PVS form ($C_\mathrm{max}=10$), and
 (f) with PMS form ($C_\mathrm{max}=10$).
 Details of the test problem are discussed in Sec. \ref{subsec:rt01}.
 \label{fig:rt01_f01}
 }
\end{figure*}

\begin{figure}[!t]
 \centering
 \includegraphics[width=0.8\hsize]{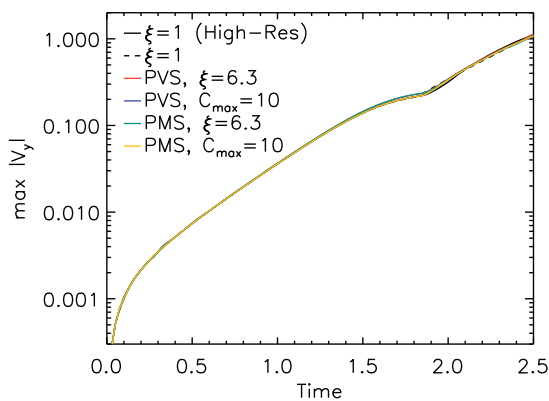}
 \caption{
 Time evolution of the Rayleigh-Taylor instability.
 The figure shows the maximum amplitude
 of the the $y$-component of the velocity without RSST ($\xi=1$)
 in $4096\times4096$ grids (black solid),
 without RSST ($\xi=1$)
 in the default $1024\times1024$ grids (black dashed),
 with PVS form ($\xi=6.3$, red),
 with PVS form ($C_\mathrm{max}=10$, blue),
 with PMS form ($\xi=6.3$, green), and
 with PMS form ($C_\mathrm{max}=10$, yellow).
 \label{fig:rt01_f02}
 }
\end{figure}

\begin{figure}[!t]
 \includegraphics[width=\hsize]{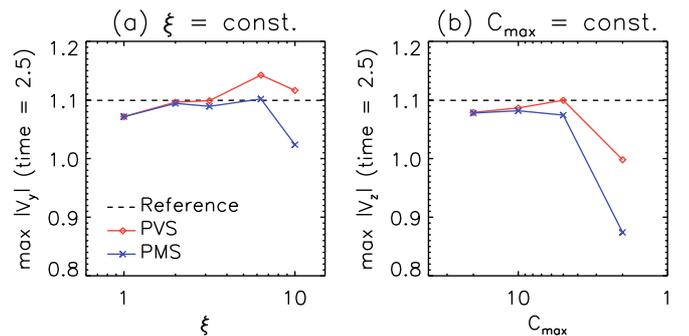}
 \caption{
 Speed of sound reduction rate
 and its dependence on the Rayleigh-Taylor instability
 due to the RSST with (a) $\xi=\mathrm{const.}$
 and (b) $C_\mathrm{max}=\mathrm{const}$.
 Each panel shows the maximum amplitude of
 the $y$-component of the velocity at time $=2.5$
 with PVS form (red with diamond) and
 PMS form (blue with cross).
 The horizontal dashed line 
 shows the reference simulation results without the RSST ($\xi=1$)
 in $4096\times4096$ grids.
 \label{fig:rt01_f04}
 }
\end{figure}

\begin{figure}[!t]
 \centering
 \includegraphics[width=0.8\hsize]{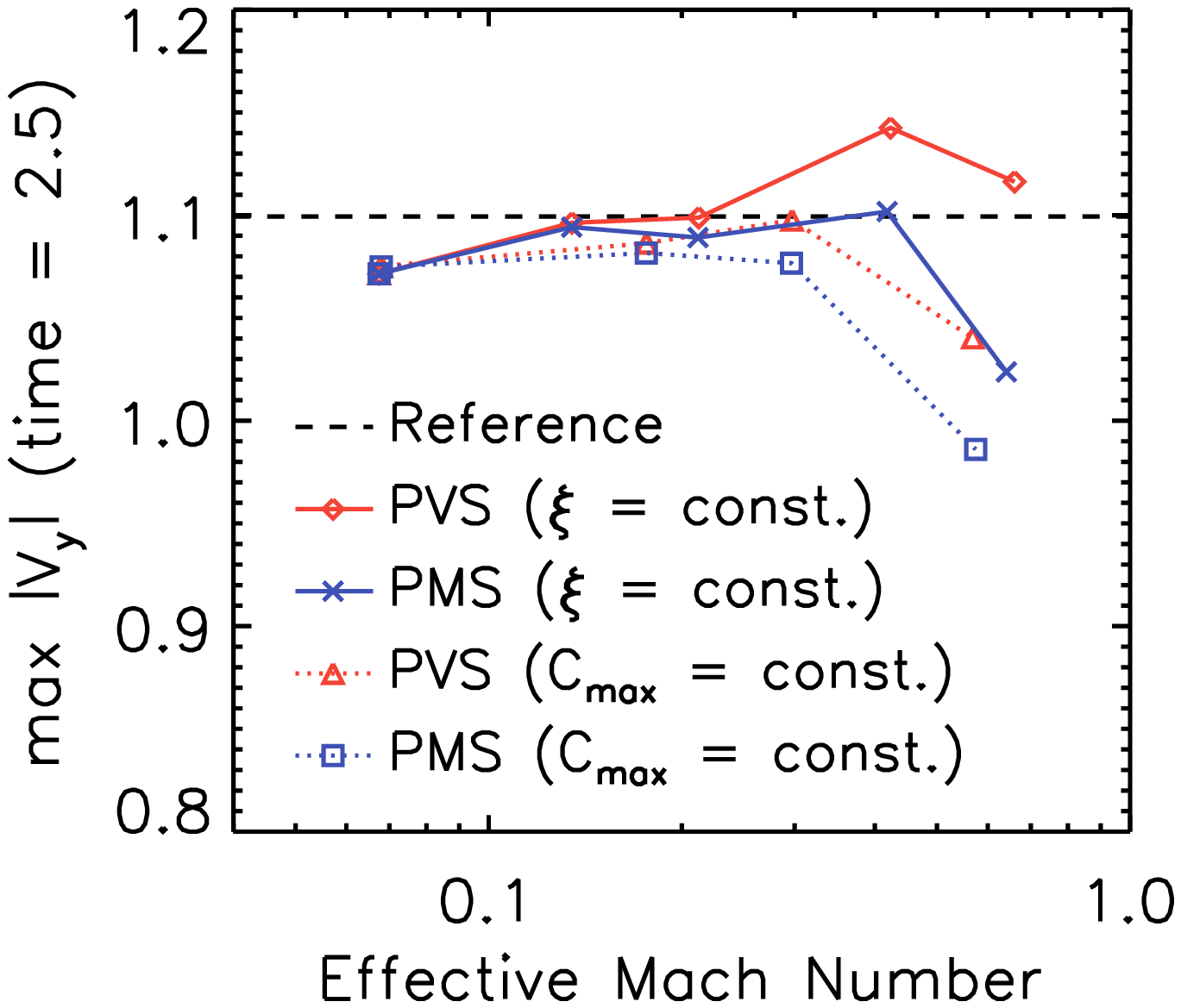}
 \caption{
 The Rayleigh-Taylor instability and its dependence on the effective Mach number.
 The maximum amplitude of
 the $y$-component of the velocity at time $=2.5$
 in PVS form with constant $\xi$
 (red solid with diamond),
 in PMS form with constant
 $\xi$ (blue solid with cross),
 in PVS form with constant $C_\mathrm{max}$
 (red dotted with triangle), and
 in PMS form with constant $C_\mathrm{max}$
 (blue dotted with box).
The horizontal dashed line 
 shows the reference simulation results without the RSST ($\xi=1$)
 in $4096\times4096$ grids.
 \label{fig:rt01_f05}
 }
\end{figure}

\subsection{Summary of the RSST variants\label{subsec:basic_eqs}}

Here, we summarize the newly suggested formulations of the RSST.
In Sec. \ref{sec:dvs} and \ref{sec:pvs},
we introduced the original DVS form of the RSST
in non-conservative form
and a new PVS form in semi-conservative form, respectively.
The DVS can be also rewritten in a semi-conservative form,
which is presented in the Appendix \ref{subadx:dvs}.
We suggest two different formulations
(PMS and DMS forms, where M stands for momentum)
that employ the fully conservative momentum equations
in Appendix \ref{subadx:pms} and \ref{subadx:dms}.
\tred{
We also construct another formulations
(PD form; Appendix \ref{subadx:pd})
that uses the strict conservative forms
in both mass and momentum equations.
We note that the PD form does not satisfy
the original form of the entropy equation.
}

In the test problem,
\tred{
we mainly focus on the PVS and PMS forms
}
because they are based on a new idea
to reduce the time evolution of the gas pressure.
\tred{
Although the PD form is also based
the idea of reducing the pressure variation,
we do not focus on this form because
it does not satisfy the entropy equation.
However, in case of the required accuracy
of the mass conservation is very severe,
the PD form might perform well.
The advantages and drawbacks of the PD form
is described in Appendix \ref{subadx:pd}.
}
We note here that the semi-conservative forms of the DVS and DMS forms
perform well in problems with small density variations,
although we do not show the results explicitly in the current paper.

\subsection{Numerical method\label{subsec:num_method}}

The proposed equations in the semi-conservative form
can be solved by
the second-order MUSCL method \citep{1979JCoPh..32..101V}.
The monotonized-central limiter is used as a slope limiter.
Temporal integration is carried out
with the second-order strong stability
preserving (SSP) Runge-Kutta method \citep{gottlieb2009high}.
The local Lax-Friedrichs (LLF) scheme is used
to compute the upwind numerical flux at the cell face.
\tred{
A Courant number of $0.4$ is used for all test problems.
}

Several modifications are required to simulate
the equations with the RSST.
The characteristic speed used in the LLF scheme
and the time step criterion must be changed to
the maximum of the absolute eigenvalue
of the corresponding RSST equations.
In this study, we used an approximate characteristic speed
$c_\mathrm{tot}=\left|\bm{V}\right|+{a}/{\xi}$
for the LLF scheme, and the time step was computed
for all proposed RSST forms.
After the time derivatives are computed
for each Runge-Kutta sub-step,
the time derivatives are modified
according to Eqs. (\ref{eq:pvs_dt0})--(\ref{eq:pvs_dt_dpr})
in the case of the PVS form.
The RSST in other forms can be computed in a similar manner.
Thus, the additional RSST computation is fully local
and incurs only minor increases in numerical computational cost.

\tred{
Here, we summarize the implementation
of the RSST (PVS form) in this study.
\begin{enumerate}
 \item Determine the size of time step using
       the Courant-Friedrich-Lewy (CFL) stability condition.
       Instead of using the maximum absolute value of exact wave speeds
       (Eqs. (\ref{eq:dvs_lambda1}) and (\ref{eq:dvs_lambda2})),
       an approximate characteristic speed
       $c_\mathrm{tot}=\left|\bm{V}\right|+{a}/{\xi}$
       is used for simplicity.
 \item Calculate limited slopes of the conservative variables.
 \item Obtain left and right states at the cell interfaces
       based on the piecewise linear reconstruction.
 \item Calculate numerical fluxes at the cell interfaces
       of the right-hand-side of Eqs. (\ref{eq:pvs_dt0}).
       We use the LLF scheme with a simplified characteristic speed
       $c_\mathrm{tot}=\left|\bm{V}\right|+{a}/{\xi}$.
 \item Obtain divergence of numerical fluxes
       to get $\varDelta{\rho}$, $\varDelta\left(\rho{V_i}\right)$,
       and $\varDelta{E}$ in Eqs. (\ref{eq:pvs_dt0}).
 \item Calculate $\varDelta{P}$ using Eq. (\ref{eq:pvs_dt_dpr}).
 \item Calculate the time derivatives of conservative variables
       using Eqs. (\ref{eq:pvs_dt}).
 \item Update the conservative variables by the Runge-Kutta method.
\end{enumerate}
}

We note a more sophisticated way
to implement the RSST in the upwind methods
by using non-conservative variants
of the path-conservative Riemann solver
\citep{dumbser2011simple,2016JCoPh.304..275D}.
Because the RSST characteristics come
from a modification of the basic equations,
we believe that the qualitative characteristics
of the results presented in this paper
are independent from the choice of numerical method.

\subsection{Functional form of the speed of sound reduction rate}

There is some freedom for the functional form
of the reduction rate of the speed of sound $\xi$.
\cite{2012A&A...539A..30H} suggest that
a spatially non-uniform reduction rate
can be also used in the RSST.
In this study, we tested two different strategies
to determine the reduction rate $\xi$.

The first strategy is to assume a spatially and temporally
constant reduction rate $\xi$.
This is easy to implement and interpret.
The time step in this case is
$\xi$-times longer than the case without the RSST,
if the Mach number is sufficiently low.
The reduction rate $\xi$ is roughly proportional
to the speed increase in the simulations.

The second choice is to assume the speed of sound has an upper limit.
Inspired by the choice of reduction rate
of the Alfv\'en speed in \cite{2009ApJ...691..640R},
we use the functional form given by
\begin{align}
 \label{eq:cx}
 \xi=\left[1+\left(a/C_\mathrm{max}\right)^4\right]^{1/4},
 \end{align}
where $C_\mathrm{max}$ is the upper limit of the speed of sound.
Eq. (\ref{eq:cx}) gives the effective speed of sound
\begin{align}
 a/\xi=a/\left[1+\left(a/C_\mathrm{max}\right)^4\right]^{1/4}.
\end{align}
The effective speed of sound $a/\xi$ approaches
$C_\mathrm{max}$ when the speed of sound $a$ increases.
The choice of Eq. (\ref{eq:cx}) can limit the speed of sound
only when reduction of the speed of sound is necessary
(i.e., when the speed of sound is similar or larger than the upper limit).

The second strategy is advantageous when the Mach number greatly varies
in the numerical domain.
For example, this applies in the solar/stellar surface convection simulation
with the deep convection zone \citep[e.g.,][]{1982A&A...107....1N}.
The numerical domain includes high Mach number shock waves
with low speed of sound near the photosphere
and low Mach number convection with high speed of sound.
In this case, the numerical time step is sometimes limited
by the high speed of sound in the deep convection zone,
and the semi-conservative RSST can
efficiently accelerate the simulation.

\tred{
\subsection{Linear Wave Convergence\label{subsec:error}}
}

\tred{
We carry out a convergence analysis
of the two-dimensional linear wave propagation
\citep[e.g.,][]{2005JCoPh.205..509G}
to verify our implementation of the RSST code.
}

\tred{
We consider the entropy wave with velocity shear propagating
at angle of $\theta=30^{\circ}$ relative to the $x$-axis.
The computational domain size
is $[0,1/\cos\theta]\times[0,1/\sin\theta]$ in the $xy$-plane.
We use ${N}\times{N}$ grid points to resolve the domain
with $N=16, 32, 64, 128, 256$.
The initial conditions are
$\rho=1+\epsilon\sin(2\pi{x_\parallel})$,
$P=1000$, $V_\parallel=1$,
$V_\perp=\epsilon\sin(2\pi{x_\parallel})$,
where $x_\parallel=x\cos\theta+y\sin\theta$ is the coordinate
along the propagation direction of linear wave
and $\epsilon=10^{-5}$ is the wave amplitude.
The subscripts $\parallel$ and $\perp$ represent
the vector components parallel and orthogonal
to the direction of wave propagation, respectively.
The velocity components along the $x$ and $y$ axes are given by
$V_x={V_\parallel}\cos\theta-{V_\perp}\sin\theta$ and
$V_y={V_\parallel}\sin\theta+{V_\perp}\cos\theta$, respectively.
The specific heat ratio of $5/3$ is used.
The initial state is evolved for unit time (until $t=1$).
We assume a spatially and temporally
uniform reduction rate $\xi=5$ in this problem.
}

\tred{
The L1 errors of the (normalized) specific entropy
$S=\ln{P}-\gamma\ln{\rho}$,
$V_\perp$, $\rho$, and $P$ for each form of the RSST
are shown in Fig. \ref{fig:err_cmp4}.
The L1 error of each variable is defined as
an volume averaged absolute value
of the difference between the initial and final snapshots.
}

\tred{
The entropy $S$ and shear velocity $V_\perp$
(panels a and b in Fig. \ref{fig:err_cmp4})
exhibit second-order convergence in all forms of the RSST
described in this paper.
On the other hand, the gas pressure $P$ (panel d)
fails to converge in the DVS and DMS forms
where the density variation is reduced.
This is caused by the pressure variation
caused by the advection of the non-uniform density (or entropy)
as discussed in Sec. \ref{subsec:dvs_drawback}.
With DVS and DMS forms,
the error of the mass density $\rho$ (panels c) shows
weak convergence with the low resolution ($N\le{32}$)
but stagnates with higher resolution.
When the pressure variation is reduced (PVS, PMS, and PD forms),
the RSST exhibits second-order convergence with $\rho$ and $P$.
}
\tred{
We note here that, in more practical problems
where sound waves are produced in the numerical domain,
even the PVS, PMS, and PD forms will fail the convergence,
because the idea of reducing the speed of sound itself
contains a source of error.
However, we believe that the error from the basic idea
can be remains in a tolerant level
in many practical problems as shown in Sec. \ref{subsec:kh01}
and \ref{subsec:rt01},
and previous studies \citep[e.g.,][]{2012A&A...539A..30H}.
}

\tred{
The above result suggests that our numerical implementation of the RSST
can reproduce the characteristics of the designed equations.
The second-order convergence is achieved
when the evolution equations of the variables
are not altered by the RSST (e.g., entropy and shear velocity).
In the following subsections,
we apply the newly suggested forms of RSST
for more complex and practical problems.
}

\subsection{Two-dimensional Kelvin-Helmholtz
  instability\label{subsec:kh01}}

The two-dimensional Kelvin-Helmholtz instability
in the low Mach number regime
can reveal the applicability of our new RSST
to complex flow structures in nearly uniform gas pressure.
We employ a version of the problem
suggested by \cite{2012ApJS..201...18M}.
The ideal equation of state was used
with the specific heat ratio of $5/3$ in all runs.
Each run is computed with a resolution of $1024\times1024$ grids.
We also carry out a high-resolution run
with $4096\times4096$ grids without applying the RSST
as a reference solution if necessary.

First, we investigate the Kelvin-Helmholtz instability in terms of
the dependence on the initial density contrast $\rho_2/\rho_1$
to clarify the difference between the newly suggested PVS/PMS forms
and the original DVS form (and a similar DMS form).
We set $\rho_1=1.0$, and the gas pressure was set to $250$
as a test problem for low Mach number flows.
The speed of sound reduction rate $\xi$ was fixed to $3$.
A typical Mach number without the RSST
is about $0.04$ in this problem.

One of the advantages of our new method based on
the reduction of pressure evolution (PVS/PMS forms)
over the original DVS form is
its applicability to flows with high density contrast.
Fig. \ref{fig:kh02_f01} demonstrates
how the PVS and DVS forms work in the Kelvin-Helmholtz instability
with different density contrasts.
We compare the cases with different density contrasts
of $\rho_2/\rho_1=1.5$ and $1.001$.
When the density contrast is small ($\rho_2/\rho_1=1.001$; bottom row),
both PMS and DVS forms work well.
However, with larger density contrast
of $\rho_2/\rho_1=1.5$ (top row),
the DVS form does not reproduce the vortex structure.
The result demonstrates that the original DVS form
can handle low Mach number flows
only when the density variation is sufficiently small.
This result is expected from the discussion
in Sec. \ref{subsec:dvs_drawback}.
The PVS form succeeds reproducing the correct time evolution,
both with low and high density contrast.
Although it is not shown in the plot,
the results with the DMS form also demonstrate
the a similar drawback as the DVS form.
The PMS form performs very similar to the PVS form.
These results clarify the advantage
of reducing the time evolution of the gas pressure in the RSST.


Next, we focus on the newly suggested PVS/PMS forms
and investigate the dependence on the reduction rate
of the speed of sound.
We set $\rho_1=1.0,\ \rho_2=2.0$, and
the gas pressure was set to $1000$.
A typical Mach number without the RSST
is about $0.02$ in this case.

Fig. \ref{fig:kh01_f01} shows
the density structure of the Kelvin-Helmholtz instability at time $=1.5$with density contrast of $2$.
The spatial structure of the density
without the RSST (panels a and d)
is successfully reproduced by PVS form (panels c and f)
and the PMS form (panels c and f).
Both reduction rate choices
($\xi$-constant or $C_\mathrm{max}$-constant)
exhibit nearly identical density.

We note that the sharpness of the vortex
is enhanced by using the RSST in Fig. \ref{fig:kh01_f01}
and is rather similar to the high-resolution run (panel d).
This is caused by reduction of the characteristic speed,
which is proportional to the numerical diffusion
in the local Lax-Friedrichs scheme used in this study.
The combination of RSST with the LLF scheme
is an efficient choice for high-resolution simulation
of low Mach number flow.

The RSST in the PVS and PMS forms reproduces the time evolution,
which is  very similar to the cases without RSST (Fig. \ref{fig:kh01_f02}).
The time evolution is characterized
by the maximum amplitude of the velocity in the $y$-direction.
This value ($\max\left|V_y\right|$) is sensitive to details
of the flow structure
compared with averaged quantities
like the root-mean-square of the velocity field.
The good agreement between the results with and without the RSST
indicates that the proposed method
maintains details of the flow structure
during evolution.
We note that there is a small periodic perturbation
in the initial phase (time $<\ 0.6$).
This perturbation is caused by
slow propagation and reflection of a sound wave
with the reduced slow speed of sound,
which disappears with smaller reduction rate
like $\xi=6.3$ or $C_\mathrm{max}=10$.

Fig. \ref{fig:kh01_f04} shows the dependence
of the maximum $y$-component velocity
on the reduction parameters ($\xi$ or $C_\mathrm{max}$)
of the speed of sound.
In the constant $\xi$ cases (panel a),
both equations (PVS and PMS forms)
shows similar $\xi$-dependence.
When the reduction rate is moderate,
the result approaches the reference solution
because the speed of sound reduction
also reduces numerical diffusion.
When the reduction rate becomes too large
and the effective Mach number approaches unity,
the error caused by the RSST increases.
Similar parameter dependence
also occurs in the constant $C_\mathrm{max}$ cases (panel b).

The dependence on the reduction parameters
in Fig. \ref{fig:kh01_f04} can be
easily interpreted by using
the effective Mach number $\xi{V}/a$.
The dependence on the effective Mach number
in Fig. \ref{fig:kh01_f05} suggests that
the effective Mach number
should be smaller than $0.3$
so that the new RSST methods
correctly reproduce the temporal evolution
of the Kelvin-Helmholtz instability.
We also note that the PVS form performs
slightly better than the PMS form in this problem,
although the both methods reproduce successful results
when the effective Mach number is sufficiently low.

\subsection{Two-dimensional Rayleigh-Taylor
  instability\label{subsec:rt01}}

We carried out a test problem for the Rayleigh-Taylor instability
to clarify the applicability of the proposed methods
under the pressure gradient and external forces like gravity.
The basic equations are extended to include the effect
of gravity $\bm{F}=(0,g_y,0)$,
as described in Appendix \ref{adx:ext_force}.
The domain size is $[0,1]\times[0,1]$ in the $xy$-plane.
The boundary condition is periodic in the $x$-direction.
The closed free-slip boundary is used in the $y$-direction.
The initial condition for density is given by
\begin{align}
 \rho=\left\{\rho_1+\frac{\rho_2-\rho_1}{2}\left[
 1+\tanh\left(\frac{y-1/2}{L}\right)\right]\right\}
 \left(1+\epsilon\right)
\end{align}
with $\rho_1=1.0$, $\rho_2=10.0$, and $L=0.025$.
$\epsilon$ is a fraction of the small perturbation
on the mass density, which is given by
\begin{align}
 \epsilon=0.01\sin\left(4\pi{x}\right)
\end{align}
The gas pressure was chosen to achieve hydrostatic balance
and is given by
\begin{align}
 P=P_c-{m_c}{g_y}
\end{align}
where the background column mass density $m_c$
from an arbitrary height $y$ to the top boundary ($y=1$)
with $\epsilon=0$ is given by
\begin{equation}
 \begin{aligned}
  m_c&=\frac{1}{2}\left(\rho_1+\rho_2\right)
  \left(1-y\right)\\
  &\hspace{0.5em}+\frac{L}{2}\left(\rho_2-\rho_1\right)
  \left[\ln\left\{
  \exp\left(\frac{1}{2L}\right)
  +\exp\left(-\frac{1}{2L}\right)
  \right\}\right.\\
  &\hspace{0.5em}\left.-\ln\left\{
  \exp\left(\frac{y-1/2}{L}\right)
  +\exp\left(-\frac{y-1/2}{L}\right)
  \right\}
  \right]
 \end{aligned}
\end{equation}
with $P_c=1000$ and gravitational acceleration $g_y=-1.0$.
All components of the initial velocity field were zero.
The ideal equation of state was used.
The specific heat ratio was set to $5/3$.
Each run was computed with a resolution of $1024\times1024$ grids.
The reference solution was simulated
in $4096\times4096$ grids without the RSST.
A typical Mach number without the RSST
is about $0.07$ in this problem.

Fig. \ref{fig:rt01_f01} shows the snapshots
from the Rayleigh-Taylor instability test problem.
The proposed RSST formulations (PVS and PMS forms)
reproduce the density pattern,
very similar to the reference case.
The functional form of the reduction rate
($\xi$ or $C_\mathrm{max}$) does not
create any clear discrepancy.

The applicability of our method
to the Rayleigh-Taylor instability can be also confirmed
from the time evolution of the maximum amplitude
of the $y$-component of the velocity (Fig. \ref{fig:rt01_f02}).
All runs shown in Fig. \ref{fig:rt01_f01}
exhibit very similar time evolution.

The dependence on the speed of sound reduction rate
is shown in Fig. \ref{fig:rt01_f04}.
As observed in the Kelvin-Helmholtz instability,
the proposed method performs well
when the reduction rate is not too high.
The threshold value of the effective Mach number
is again $0.3$ (Fig. \ref{fig:rt01_f05}),
as was the case in the test problem for the Kelvin-Helmholtz instability.

\section{Summary and discussion}

In this study, we proposed several new formulations
of the reduced speed of sound technique (RSST),
which has been applied to accelerate the computational speed
in simulations of low Mach number flows.
\tred{
The convergence test of the linear entropy wave
and more practical Kelvin-Helmholtz
and Rayleigh-Taylor instability problems are carried out.
}
The numerical tests suggest that
the effective Mach number (after reducing the speed of sound)
should be less than $0.3$
in order to maintain the characteristics of the flows.
We note that the methods can be easily
extended to the magnetohydrodynamic equations.

\tred{
We note that all of new formulations of the RSST are derived
by assuming the general equation of state for non-ideal gas.
Although all of the test problems presented in this paper
assumes the ideal equation of state.
we have also carried out several tests
using the van der Waals equation of state
\citep[e.g., ][]{2014IJNMF..75..467C}.
Because we could not find any qualitative difference
from the ideal equation of state,
we believe that our method can work even
with a general equation of state.
}

We summarize the characteristic points of the proposed methods.
All of the RSST described in this paper
(DVS, DMS, PVS, \tred{PMS, and PD})
share several characteristics:
\begin{itemize}
 \item The methods can be easily implemented in an explicit scheme
       and can accelerate the computational speed
       of low Mach number flows.
 \item The steady solution of the RSST equations
       is the same as the steady solution of the original Euler equations.
 \item The methods are formulated in semi-conservative form
       so that the equations reduce to the conservative Euler equations
       when the RSST is not used.
\end{itemize}
The subgroups of the proposed methods
have the following characteristics:
\begin{itemize}
 \item The newly proposed methods based on the reduction
       of the temporal evolution of gas pressure
       (PVS, \tred{PMS, and PD} forms)
       share the advantage that the proposed methods can be applied
       to flows with large density variation.
 \item The methods based on reduction of density variation
       (DVS and DMS forms) preserve the modified mass conservation law
       ($\partial\left<\xi\rho\right>/\partial{t}=0$,
       where $\left<.\right>$ indicates a volume average)
       if the reduction rate is time independent
       ($\partial\xi/\partial{t}=0$).
 \item The DMS and PMS methods employ the exact conservative form
       of the momentum equations so that the volume integral of the momentum
       can be conserved down to the round-off error
       through combination with the finite volume or finite difference methods.
       \tred{
 \item The PD form is based on the exact conservation laws
       of both mass and momentum.
       However, the correct evolution of the entropy
       can be violated by the pressure variation
       (see also Appendix \ref{subadx:pd}).
       }
\end{itemize}
These various characteristics will broaden
the application range of the RSST
to a variety of low Mach number phenomena.



\begin{acknowledgements}
This work is supported by MEXT/JSPS KAKENHI Grant Number 15H05816.
This work was carried out by the joint research program of the Institute for
Space-Earth Environment Research (ISEE), Nagoya University.
Numerical computations were carried out
on a Cray XC50 supercomputer at the Center
for Computational Astrophysics, National Astronomical Observatory of Japan.
\end{acknowledgements}

\appendix

\section{Variants of semi-conservative RSST\label{adx:rsst}}

In this Appendix,
we describe the original and \tred{three} alternatives of RSST
(DVS, DMS, \tred{PMS, and PD} forms) in their semi-conservative forms.
The two new RSST formulations (DMS and PMS forms)
are based on the conservative form of the momentum equations
rather than the primitive equations of motion,
as was the case in the DVS and PVS forms.
Accurate conservation of the momentum
will be favored in some situations.
The PMS form described in Appendix \ref{subadx:pms}
is based on the reduced pressure evolution.
On the other hand, the DMS form described in Appendix \ref{subadx:dms}
is similar to the DVS form and is derived
by reducing the density evolution.
\tred{
The PD form reduces the pressure variation as in the PVS and PMS forms.
This new form has the superior conservative property
of both mass and momentum in the system,
but it does not satisfy the strict form of the entropy equation.
}

\subsection{DVS form for conservative variables\label{subadx:dvs}}

The original DVS form can be rewritten
in a semi-conservative form given by
\begin{equation}
  \begin{aligned}
   \frac{\partial{\rho}}{\partial{t}}&
   +\nabla\cdot\left(\rho\bm{V}\right)=
   -\left(1-\frac{1}{\xi^2}\right)\varDelta{\rho}
   \\
   \frac{\partial}{\partial{t}}\left({\rho{V_i}}\right)&
   +\nabla\cdot\left(\rho{V_i}\bm{V}\right)
   +\frac{\partial{P}}{\partial{x_i}}=
   -\left(1-\frac{1}{\xi^2}\right)
   \left(
   \frac{\partial{\rho{V_i}}}{\partial{\rho}}
   \right)_{\bm{V}}\varDelta{\rho}
   \\
   \frac{\partial{E}}{\partial{t}}&
   +\nabla\cdot\left[\left(E+P\right)\bm{V}\right]
   =
   -\left(1-\frac{1}{\xi^2}\right)
   \left(
   \frac{\partial{E}}{\partial{\rho}}
   \right)_{\bm{V},S}\varDelta{\rho}
  \end{aligned}
\end{equation}
where
\begin{align}
 \varDelta{\rho}=-\nabla\cdot\left(\rho\bm{V}\right)
\end{align}
is the time variation of the mass density without the RSST
and
\begin{align}
 \left(
 \frac{\partial{\rho{V_i}}}{\partial{\rho}}
 \right)_{\bm{V}}
 =V_i
 ,\
 \left(
 \frac{\partial{E}}{\partial{\rho}}
 \right)_{\bm{V},S}
 =\frac{e+P+\rho{V^2}/2}{\rho}
\end{align}
As shown in Sec. \ref{sec:test},
the DVS form in this semi-conservative formulation
also has a drawback in that it is not applicable
to flows with large density variation.

\subsection{Momentum conservative form based on
  pressure variation reduction (PMS form)\label{subadx:pms}}

The PMS form is an alternative of the PVS form (Sec. \ref{sec:pvs})
that strictly conserves the momentum of the system
if the momentum flux through the boundary is negligible.
The basic equations of the PMS form are
the evolution equations of $(P,\rho\bm{V},S)$, which are given by
\begin{equation}
 \label{eq:pms_prim}
  \begin{aligned}
 \xi^2\frac{\partial{P}}{\partial{t}}
 +\bm{V}\cdot\nabla{P}
 +\rho{a^2}\nabla\cdot\bm{V}=0\\
 \frac{\partial}{\partial{t}}\left({\rho{V_i}}\right)
 +\nabla\cdot\left(\rho{V_i}\bm{V}\right)
 +\frac{\partial{P}}{\partial{x_i}}=0\\
 \frac{\partial{S}}{\partial{t}}+\bm{V}\cdot\nabla{S}=0
  \end{aligned}
\end{equation}
The only difference from the PVS form is
the use of the conservative form of the momentum equations
instead of the primitive equations of motion.
Apparently, this formulation conserves
the volume average of the momentum in the isolated system.

The phase speeds of each wave mode
in Eqs. (\ref{eq:pms_prim}) are different from those
in the DVS or PVS forms.
From the one-dimensional version of Eqs. (\ref{eq:pms_prim}),
\begin{align}
 &\frac{\partial}{\partial{t}}
 \begin{pmatrix}
  P\\
  \rho{V_x}\\
  S
 \end{pmatrix}
 +A\frac{\partial}{\partial{x}}
 \begin{pmatrix}
  P\\
  \rho{V_x}\\
  S
 \end{pmatrix}
 =0\\
 A&=
 \begin{pmatrix}
  0 & a^2/\xi^2 & P_SV_x/\xi^2\\
  1-V_x^2/a^2 & 2V_x & P_S{V_x^2}/a^2\\
  0 & 0 & V_x\\
 \end{pmatrix}
\end{align}
the wave speeds are given by
\begin{align}
 \label{eq:pms_speed}
 \lambda=
 \begin{cases}
  V_x,\\
  V_x\pm\sqrt{D}
 \end{cases}\ \mathrm{where}\
 D=\left(1-\frac{1}{\xi^2}\right)V_x^2+\frac{1}{\xi^2}a^2
\end{align}
Because $\xi\ge1$ is always satisfied in order to limit the speed of sound,
all wave speeds are real ($D>0$)
and Eqs. (\ref{eq:pms_prim}) are hyperbolic.
The effective speed of sound $\sqrt{D}$
is larger than the absolute value of the velocity $|V_x|$
and is smaller than the real speed of sound $a$.
In the low Mach number limit ($|V_x|\ll{a}$),
the effective speed of sound $\sqrt{D}$ approaches $a/\xi$
as was the case in the original RSST.
Thus, this new formulation can be used
to reduce the speed of sound in low Mach number flows.

The momentum conserving PMS form
can be also rewritten as evolution equations
in terms of the conservative variables.
The semi-conservative form of Eqs. (\ref{eq:pms_prim}) is given by
\begin{equation}
  \begin{aligned}
 \frac{\partial{\rho}}{\partial{t}}&
 +\nabla\cdot\left(\rho\bm{V}\right)=
 -\left(1-\frac{1}{\xi^2}\right)
 \left(
 \frac{\partial{\rho}}{\partial{P}}
 \right)_{S}\varDelta{P}
 \\
 \frac{\partial}{\partial{t}}\left({\rho{V_i}}\right)&
 +\nabla\cdot\left(\rho{V_i}\bm{V}\right)
 +\frac{\partial{P}}{\partial{x_i}}=0
 \\
 \frac{\partial{E}}{\partial{t}}&
 +\nabla\cdot\left[\left(E+P\right)\bm{V}\right]
 =
 -\left(1-\frac{1}{\xi^2}\right)
 \left(
 \frac{\partial{E}}{\partial{P}}
 \right)_{\rho\bm{V},S}\varDelta{P}
  \end{aligned}
\end{equation}
where $\varDelta{P}$ has the same
definition in Eq. (\ref{eq:pvs_cons_dpr}),
and
\begin{align}
 \left(
 \frac{\partial{E}}{\partial{P}}
 \right)_{\rho\bm{V},S}
 =\frac{e+P-\rho{V^2}/2}{\rho{a^2}}
\end{align}
Apparently, the momentum equations are
in a complete conservative form.

As tested in Sec. \ref{sec:test},
we cannot find any significant drawback
that arises from using the conservative form in the momentum equations.
For application to problems where
 (angular) momentum conservation is important,
the PMS form will have some advantages.

\subsection{Momentum conservative form based on
  density variation reduction (DMS form)\label{subadx:dms}}

The DMS form is based on reducing the density variation,
similar to the DVS form
(Sec. \ref{sec:dvs} and Appendix \ref{subadx:dvs}).
The difference is that the DMS form employs
the conservative form of the momentum equations.
The basic equations of the DMS form are
the evolution equations of $(\rho,\rho\bm{V},S)$, which are given by
\begin{equation}
 \label{eq:dms_prim}
  \begin{aligned}
   \xi^2\frac{\partial{\rho}}{\partial{t}}
   +\nabla\cdot\left(\rho\bm{V}\right)=0\\
   \frac{\partial}{\partial{t}}\left({\rho{V_i}}\right)
   +\nabla\cdot\left(\rho{V_i}\bm{V}\right)
   +\frac{\partial{P}}{\partial{x_i}}=0\\
   \frac{\partial{S}}{\partial{t}}+\bm{V}\cdot\nabla{S}=0
  \end{aligned}
\end{equation}
The wave speed is same as the PMS form
and is given in Eq. (\ref{eq:pms_speed}).

The semi-conservative version of the DMS form
is given by
\begin{equation}
  \begin{aligned}
 \frac{\partial{\rho}}{\partial{t}}&
 +\nabla\cdot\left(\rho\bm{V}\right)=
 -\left(1-\frac{1}{\xi^2}\right)\varDelta{\rho}
 \\
 \frac{\partial}{\partial{t}}\left({\rho{V_i}}\right)&
 +\nabla\cdot\left(\rho{V_i}\bm{V}\right)
 +\frac{\partial{P}}{\partial{x_i}}=0
 \\
 \frac{\partial{E}}{\partial{t}}&
 +\nabla\cdot\left[\left(E+P\right)\bm{V}\right]
 =
 -\left(1-\frac{1}{\xi^2}\right)
 \left(
 \frac{\partial{E}}{\partial{\rho}}
 \right)_{\rho\bm{V},S}\varDelta{\rho}
  \end{aligned}
\end{equation}
where $\varDelta{\rho}$ has the same
definition in Eq. (\ref{eq:pvs_cons_dpr}),
and
\begin{align}
 \left(
 \frac{\partial{E}}{\partial{\rho}}
 \right)_{\rho\bm{V},S}
 =\frac{e+P-\rho{V^2}/2}{\rho}
\end{align}
Because the drawback of the DVS form
described in Sec. \ref{subsec:dvs_drawback}
is independent from the form of
the momentum equations in the basic equations,
the DMS form has the same drawback
as was the case in the DVS form.

\tred{
\subsection{Mass and momentum conservative form based on
  pressure variation reduction (PD form)\label{subadx:pd}}
}

\tred{
We can also construct a form of the RSST
that strictly conserves both mass and momentum in the closed system.
The PD form is based on the pressure variation reduction method.
\begin{equation}
 \label{eq:pd_prim}
  \begin{aligned}
   \xi^2\frac{\partial{P}}{\partial{t}}
   +\bm{V}\cdot\nabla{P}
   +\rho{a^2}\nabla\cdot\bm{V}=0\\
   \frac{\partial{V_i}}{\partial{t}}+\bm{V}\cdot\nabla{V_i}
   +\frac{1}{\rho}\frac{\partial{P}}{\partial{x_i}}=0\\
   \frac{\partial{\rho}}{\partial{t}}
   +\nabla\cdot\left(\rho\bm{V}\right)=0
  \end{aligned}
\end{equation}
The wave speeds $\lambda$ of Eqs. (\ref{eq:pd_prim})
are identical to the speeds in DVS and PVS forms and given by
Eqs. (\ref{eq:dvs_lambda1}) and (\ref{eq:dvs_lambda2}).
}

\tred{
The semi-conservative equations of the PD form is given by
\begin{equation}
  \begin{aligned}
   \frac{\partial{\rho}}{\partial{t}}&
   +\nabla\cdot\left(\rho\bm{V}\right)=0
   \\
   \frac{\partial}{\partial{t}}\left({\rho{V_i}}\right)&
   +\nabla\cdot\left(\rho{V_i}\bm{V}\right)
   +\frac{\partial{P}}{\partial{x_i}}=0
   \\
   \frac{\partial{E}}{\partial{t}}&
   +\nabla\cdot\left[\left(E+P\right)\bm{V}\right]
   =
   -\left(1-\frac{1}{\xi^2}\right)
   \left(
   \frac{\partial{E}}{\partial{P}}
   \right)_{\bm{V},\rho}\varDelta{P}
  \end{aligned}
\end{equation}
where
\begin{align}
 \left(\frac{\partial{E}}{\partial{P}}\right)_{\bm{V},\rho}
 =\left(\frac{\partial{e}}{\partial{P}}\right)_{\rho}
\end{align}
and the $\varDelta{P}$ is defined by Eq. (\ref{eq:pvs_cons_dpr}).
Apparently, the PD form employs the exact form of
the mass and momentum conservation laws.
This characteristic is advantageous when both mass
and momentum conservation is important.
}

\tred{
Although the PD form has the superior conservative property
of the mass and momentum,
the user should be careful when this form is applied
to the practical problems.
From Eqs. (\ref{eq:pd_prim}),
the entropy equation in the PD form is given by
\begin{align}
 \label{eq:pd_entropy}
 P_S\frac{DS}{Dt}=\left(1-\frac{1}{\xi^2}\right)\varDelta{P}
\end{align}
where $P_S=(\partial{P}/\partial{S})_\rho$
as described in Sec. \ref{subsec:dvs_prim}.
Eq. (\ref{eq:pd_entropy}) indicates that the pressure variation
(e.g., sound wave) can change the specific entropy artificially.
Although the PD form performs well
in all of the test problems described in this paper,
such violation of the entropy evolution
can be a possible source of error in more severe problems.
One example is the thermal convection
in the deep stellar convection zone
where the very small variation of the entropy
drives the convective motion through the buoyancy force
(i.e., the super-adiabaticity is very small),
although the convective motion continuously excites sound waves.
}

\section{Extension of the RSST with an external force\label{adx:ext_force}}

We need to include the effect of an external force (gravity)
for simulation of the two-dimensional
Rayleigh-Taylor instability in Sec. \ref{subsec:rt01}.

The basic equations of the PVS form are extended
to the case with an external force $\bm{F}$ as follows:
(1) We add $\rho\bm{F}$ and $\rho\bm{V}\cdot\bm{F}$
to the right hand side of the momentum and total energy equations
of Eqs. (\ref{eq:pvs_cons}) or Eqs. (\ref{eq:pvs_dt0}),
respectively.
(2) The definition of $\varDelta{P}$
in Eq. (\ref{eq:pvs_cons_dpr}) or (\ref{eq:pvs_dt_dpr})
remains unchanged.
The resulting equations are given by
\begin{equation}
 \begin{aligned}
  \frac{\partial{\rho}}{\partial{t}}&
  +\nabla\cdot\left(\rho\bm{V}\right)=
  -\left(1-\frac{1}{\xi^2}\right)
  \left(
  \frac{\partial{\rho}}{\partial{P}}
  \right)_{S}\varDelta{P}
  \\
  \frac{\partial}{\partial{t}}\left({\rho{V_i}}\right)&
  +\nabla\cdot\left(\rho{V_i}\bm{V}\right)
  +\frac{\partial{P}}{\partial{x_i}}=
  -\left(1-\frac{1}{\xi^2}\right)
  \left(
  \frac{\partial{\rho{V_i}}}{\partial{P}}
  \right)_{\bm{V},S}\varDelta{P}
  +\rho\bm{F}
  \\
  \frac{\partial{E}}{\partial{t}}&
  +\nabla\cdot\left[\left(E+P\right)\bm{V}\right]
  =
  -\left(1-\frac{1}{\xi^2}\right)
  \left(
  \frac{\partial{E}}{\partial{P}}
  \right)_{\bm{V},S}\varDelta{P}
  +\rho\bm{V}\cdot\bm{F}
 \end{aligned}
\end{equation}
where $\varDelta{P}$
and the partial derivatives $\left(\partial{\rho}/\partial{P}\right)_{S}$,
$\left(\partial{\rho{V_i}}/\partial{P}\right)_{\bm{V},S}$, and
$\left(\partial{E}/\partial{P}\right)_{\bm{V},S}$
are given in Eqs. (\ref{eq:pvs_cons_dpr})
and (\ref{eq:pvs_cons_rhs}), respectively.

The PMS, DVS, DMS, and PD forms
can be extended in a similar fashion.



\begin{thebibliography}{28}
\expandafter\ifx\csname natexlab\endcsname\relax\def\natexlab#1{#1}\fi

\bibitem[{{Cai}(2016)}]{2016JCoPh.310..342C}
{Cai}, T. 2016, Journal of Computational Physics, 310, 342

\bibitem[{{Castro} \& {Toro}(2014)}]{2014IJNMF..75..467C}
{Castro}, C.~E. \& {Toro}, E.~F. 2014, International Journal for Numerical
  Methods in Fluids, 75, 467

\bibitem[{{Chan} {et~al.}(1994){Chan}, {Mayr}, {Mengel}, \&
  {Harris}}]{1994JCoPh.113..165C}
{Chan}, K.~L., {Mayr}, H.~G., {Mengel}, J.~G., \& {Harris}, I. 1994, Journal of
  Computational Physics, 113, 165

\bibitem[{{Dumbser} \& {Balsara}(2016)}]{2016JCoPh.304..275D}
{Dumbser}, M. \& {Balsara}, D.~S. 2016, Journal of Computational Physics, 304,
  275

\bibitem[{Dumbser \& Toro(2011)}]{dumbser2011simple}
Dumbser, M. \& Toro, E.~F. 2011, Journal of Scientific Computing, 48, 70

\bibitem[{{Gardiner} \& {Stone}(2005)}]{2005JCoPh.205..509G}
{Gardiner}, T.~A. \& {Stone}, J.~M. 2005, Journal of Computational Physics,
  205, 509

\bibitem[{{Glasner} {et~al.}(2007){Glasner}, {Livne}, \&
  {Truran}}]{2007ApJ...665.1321G}
{Glasner}, S.~A., {Livne}, E., \& {Truran}, J.~W. 2007, \apj, 665, 1321

\bibitem[{{Gottlieb} {et~al.}(2009){Gottlieb}, {Ketcheson}, \&
  {Shu}}]{gottlieb2009high}
{Gottlieb}, S., {Ketcheson}, D.~I., \& {Shu}, C.-W. 2009, Journal of Scientific
  Computing, 38, 251

\bibitem[{{Hotta} {et~al.}(2014){Hotta}, {Rempel}, \&
  {Yokoyama}}]{2014ApJ...786...24H}
{Hotta}, H., {Rempel}, M., \& {Yokoyama}, T. 2014, \apj, 786, 24

\bibitem[{{Hotta} {et~al.}(2015){Hotta}, {Rempel}, \&
  {Yokoyama}}]{2015ApJ...798...51H}
{Hotta}, H., {Rempel}, M., \& {Yokoyama}, T. 2015, \apj, 798, 51

\bibitem[{{Hotta} {et~al.}(2016){Hotta}, {Rempel}, \&
  {Yokoyama}}]{2016Sci...351.1427H}
{Hotta}, H., {Rempel}, M., \& {Yokoyama}, T. 2016, Science, 351, 1427

\bibitem[{{Hotta} {et~al.}(2012){Hotta}, {Rempel}, {Yokoyama}, {Iida}, \&
  {Fan}}]{2012A&A...539A..30H}
{Hotta}, H., {Rempel}, M., {Yokoyama}, T., {Iida}, Y., \& {Fan}, Y. 2012, \aap,
  539, A30

\bibitem[{{Hou} \& {Lefloch}(1994)}]{1994MaCom..62..497H}
{Hou}, T.~Y. \& {Lefloch}, P.~G. 1994, Mathematics of Computation, 62, 497

\bibitem[{{Iijima} \& {Yokoyama}(2015)}]{2015ApJ...812L..30I}
{Iijima}, H. \& {Yokoyama}, T. 2015, \apjl, 812, L30

\bibitem[{{Iijima} \& {Yokoyama}(2017)}]{2017ApJ...848...38I}
{Iijima}, H. \& {Yokoyama}, T. 2017, \apj, 848, 38

\bibitem[{{K{\"a}pyl{\"a}} {et~al.}(2016){K{\"a}pyl{\"a}}, {Brandenburg},
  {Kleeorin}, {K{\"a}pyl{\"a}}, \& {Rogachevskii}}]{2016A&A...588A.150K}
{K{\"a}pyl{\"a}}, P.~J., {Brandenburg}, A., {Kleeorin}, N., {K{\"a}pyl{\"a}},
  M.~J., \& {Rogachevskii}, I. 2016, \aap, 588, A150

\bibitem[{{Kupka} \& {Muthsam}(2017)}]{2017LRCA....3....1K}
{Kupka}, F. \& {Muthsam}, H.~J. 2017, Living Reviews in Computational
  Astrophysics, 3, 1

\bibitem[{{McNally} {et~al.}(2012){McNally}, {Lyra}, \&
  {Passy}}]{2012ApJS..201...18M}
{McNally}, C.~P., {Lyra}, W., \& {Passy}, J.-C. 2012, \apjs, 201, 18

\bibitem[{{Miesch} {et~al.}(2008){Miesch}, {Brun}, {De Rosa}, \&
  {Toomre}}]{2008ApJ...673..557M}
{Miesch}, M.~S., {Brun}, A.~S., {De Rosa}, M.~L., \& {Toomre}, J. 2008, \apj,
  673, 557

\bibitem[{{Nonaka} {et~al.}(2010){Nonaka}, {Almgren}, {Bell}, {Lijewski},
  {Malone}, \& {Zingale}}]{2010ApJS..188..358N}
{Nonaka}, A., {Almgren}, A.~S., {Bell}, J.~B., {et~al.} 2010, \apjs, 188, 358

\bibitem[{{Nordlund}(1982)}]{1982A&A...107....1N}
{Nordlund}, {\AA}. 1982, \aap, 107, 1

\bibitem[{{Rempel}(2005)}]{2005ApJ...622.1320R}
{Rempel}, M. 2005, \apj, 622, 1320

\bibitem[{{Rempel} {et~al.}(2009){Rempel}, {Sch{\"u}ssler}, \&
  {Kn{\"o}lker}}]{2009ApJ...691..640R}
{Rempel}, M., {Sch{\"u}ssler}, M., \& {Kn{\"o}lker}, M. 2009, \apj, 691, 640

\bibitem[{{Stein} \& {Nordlund}(1989)}]{1989ApJ...342L..95S}
{Stein}, R.~F. \& {Nordlund}, {\AA}. 1989, \apjl, 342, L95

\bibitem[{{Takeyama} {et~al.}(2017){Takeyama}, {Saitoh}, \&
  {Makino}}]{2017NewA...50...82T}
{Takeyama}, K., {Saitoh}, T.~R., \& {Makino}, J. 2017, \na, 50, 82

\bibitem[{{van Leer}(1979)}]{1979JCoPh..32..101V}
{van Leer}, B. 1979, Journal of Computational Physics, 32, 101

\bibitem[{{V{\"o}gler} {et~al.}(2005){V{\"o}gler}, {Shelyag}, {Sch{\"u}ssler},
  {Cattaneo}, {Emonet}, \& {Linde}}]{2005A&A...429..335V}
{V{\"o}gler}, A., {Shelyag}, S., {Sch{\"u}ssler}, M., {et~al.} 2005, \aap, 429,
  335

\bibitem[{{Zingale} {et~al.}(2005){Zingale}, {Woosley}, {Rendleman}, {Day}, \&
  {Bell}}]{2005ApJ...632.1021Z}
{Zingale}, M., {Woosley}, S.~E., {Rendleman}, C.~A., {Day}, M.~S., \& {Bell},
  J.~B. 2005, \apj, 632, 1021

\end{thebibliography}

\end{document}